\documentclass[aps,pra,reprint,floatfix,superscriptaddress,nofootinbib]{revtex4-2}

\usepackage[T1]{fontenc}
\usepackage[utf8]{inputenc}
\usepackage{amsmath,amssymb,bm}
\usepackage{graphicx}
\usepackage[hidelinks]{hyperref}
\usepackage{booktabs}
\usepackage{microtype}
\usepackage{xcolor}
\graphicspath{{figures/}}

\newcommand{\avg}[1]{\left\langle #1 \right\rangle}
\newcommand{\Var}{\mathrm{Var}}

\begin{document}

\title{Memory-assisted squeezed light velocimetry under realistic loss and incoherent noise}

\author{Mustafa G\"undo\u{g}an}
\email{mustafa.guendogan@physik.hu-berlin.de}
\affiliation{Institut f\"ur Physik and Center for the Science of Materials Berlin (CSMB), Humboldt-Universit\"at zu Berlin, Newtonstr.~15, 12489 Berlin, Germany}

\author{Arash Ahmadi}
\affiliation{Institut f\"ur Physik and Center for the Science of Materials Berlin (CSMB), Humboldt-Universit\"at zu Berlin, Newtonstr.~15, 12489 Berlin, Germany}

\author{Markus Krutzik}
\affiliation{Institut f\"ur Physik and Center for the Science of Materials Berlin (CSMB), Humboldt-Universit\"at zu Berlin, Newtonstr.~15, 12489 Berlin, Germany}
\affiliation{Ferdinand-Braun-Institut (FBH), Gustav-Kirchhoff-Str.~4, 12489 Berlin, Germany}

\date{\today}

\begin{abstract}
We propose a velocity sensor based on a two-memory Mach--Zehnder interferometer fed by a coherent probe and squeezed vacuum, read out by balanced homodyne detection. One memory is taken as a stationary reference, while the second memory moves during storage, so that its velocity is mapped onto a differential interferometric phase at readout. The two memories are otherwise assumed identical and are described by a Gaussian write--store--read lifetime together with the associated unconditional noise floor. Using the classical Fisher information, we derive the velocity sensitivity, the transmission threshold required for a target quantum gain, and the optimum storage time. The squeezed scheme improves on equal-resource coherent homodyne within an operating window set mainly by total transmission and phase stability. For representative near-term parameters, unconditional memory noise floors up to about $10^{-1}$ photons per trial do not by themselves remove the advantage; after optimization the improvement remains at the few-percent level and is limited chiefly by loss.\end{abstract}

\maketitle

\section{Introduction}

Interferometric phase sensing remains one of the most mature settings for quantum enhancement. Injecting squeezed vacuum into the dark port of a Mach-Zehnder interferometer reduces the detected phase noise below the coherent-state limit, but the improvement survives only when attenuation and technical fluctuations remain sufficiently small~\cite{Caves1981,Giovannetti2004,Pezze2008,Ono2010,Lang2013,Demkowicz2015,Paris2009,Schnabel2017}. Optical memories~\cite{Fleischhauer2000,Fleischhauer2002,Lvovsky2009,Sangouard2011,Heshami2016,Lei2023} add another control parameter to this setting. By storing an optical pulse during a controllable period, they allow an external perturbation to accumulate as a phase of the retrieved light, at the price of storage time-dependent loss and memory-generated noise. Related ideas have also started to appear in gravitationally motivated interferometry, where quantum memories act as interrogation elements or time-delay resources in memory-assisted Mach-Zehnder, Hong-Ou-Mandel, and vertically separated clock-like settings \cite{Barzel2024,Gundogan2026Gravity}.

A particularly relevant application in this direction is velocity sensing with stored light, which has recently been proposed and experimentally explored in memory-based architectures \cite{Ahmadi2025Concept,Ahmadi2025Experiment}. The basic idea is to map an optical field into a collective excitation, let it evolve during the storage interval, and then read the light out interferometrically. Since such optical memories operates in a phase preserving way, the output light has a fixed phase in the memory frame. However, when physically translated, this corresponds to a phase shift in the lab frame.  For a probe of wave number $k_p=2\pi/\lambda_p$, with $\lambda_p$ the probe wavelength, this velocity-induced phase can be written as
\begin{equation}
\phi_v=-k_p v\tau,
\label{eq:phi_v}
\end{equation}
where $v$ is the average velocity during the storage time $\tau$. Increasing $\tau$ strengthens the transduced phase, but at the same time the retrieval efficiency of the memories decreases. Experiments on phase-preserving optical storage in solids, cold atoms, and warm vapors \cite{Turukhin2001,Heinze2013, Gundogan2013, Hsiao2018,Katz2018,Ma2022,Mair2002,Chen2005Beat,Jeong2017}, storage of squeezed states~\cite{Appel2008, Honda2008, Arikawa2010}, together with motion-sensitive slow-light and light-dragging measurements \cite{Kuan2016,Chen2020QEV,Solomons2020,Banerjee2022} and transport of stored light in moving or reconfigured media \cite{Juzeliunas2003,Zibrov2002,Ginsberg2007,Li2020}, provide the experimental background for such a device. Classical laser-Doppler and self-mixing velocimeters remain the relevant nonquantum benchmarks away from the shot-noise limit \cite{Norgia2017,Sun2018,Zhang2019}.

In the present work, we ask whether the memory-assisted velocimetry platform can be pushed beyond the coherent-state shot-noise limit by replacing classical probe light with genuinely quantum input states. More specifically, we consider whether squeezed-input interferometry can retain a metrological advantage once the full sensing cycle is taken into account: propagation through the external optics, storage and retrieval in the memories, and final homodyne readout. This is a stricter question than ideal loss-only interferometry, because in memory-based sensing the same storage period that amplifies the velocity-dependent phase also reduces the retrieved signal and introduces additional incoherent noise associated with the memory operation \cite{Gundogan2015,Heller2022,Esguerra2023,Wang2022,Jutisz2025}.

To this end, here we develop a Gaussian model for a two-memory Mach-Zehnder velocimeter operated near the dark fringe and read out by balanced homodyne detection, with squeezed vacuum injected at the input. The model includes Gaussian memory lifetime, external optical loss, detector inefficiency, interferometer phase noise, squeezing-angle jitter, electronic noise, and the unconditional memory noise floor $p_n$. Both memories are taken to be identical in their efficiency and noise properties. Within this framework, we derive the Cram\'er-Rao bound for velocity estimation, identify the minimum total transmission required for a prescribed quantum advantage over the coherent-state benchmark, and determine the optimum storage time when the memory efficiency follows a Gaussian decay law.

\section{Model}

\subsection{Interferometer configuration and velocity encoding}

\begin{figure}[t]
    \centering
    \includegraphics[width=\columnwidth]{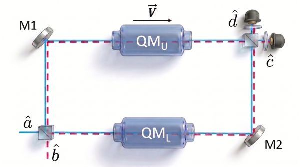}
    \caption{Two-memory Mach-Zehnder model. A coherent probe enters mode $\hat a_0$ and phase-squeezed vacuum enters mode $\hat b_0$. Both arms are stored and retrieved in identical memories with write-store-read efficiency $\eta_{\rm mem}(\tau)=\eta_0\exp[-(\tau/\tau_{\rm mem})^2]$ and the same unconditional noise floor $p_n(\tau)$. Motion of the upper memory during storage generates the signal phase $\phi_v=-k_p v\tau$. Balanced homodyne detection measures the dark output.}
    \label{fig:schematic}
\end{figure}

The sensing geometry is shown in Fig.~\ref{fig:schematic}. A bright coherent state $|\alpha\rangle$ with $N\equiv|\alpha|^2$ photons per interrogation enters input mode $\hat a_0$. A single-mode squeezed vacuum state $S(r)|0\rangle$ enters the unused input mode $\hat b_0$, where
\begin{equation}
S(r)=\exp\!\left[\frac{r}{2}\left(\hat b_0^2-\hat b_0^{\dagger 2}\right)\right]
\label{eq:squeeze_op}
\end{equation}
is the squeeze operator and $r\ge 0$ is the squeezing parameter. We use quadratures $
\hat X=\hat b+\hat b^\dagger$ and $\hat P={(\hat b-\hat b^\dagger)}/{i}$ for which vacuum has unit variance, $\Var(\hat X)=\Var(\hat P)=1$. The injected squeezed state is aligned so that the phase-relevant quadrature is squeezed at the source, $\Var(\hat P_{b_0})=\exp(-2r)$.

The first $50{:}50$ beam splitter produces arm modes
\begin{equation}
\hat u_0=\frac{\hat a_0+\hat b_0}{\sqrt{2}},
\qquad
\hat \ell_0=\frac{\hat a_0-\hat b_0}{\sqrt{2}},
\label{eq:arm_modes}
\end{equation}
where $\hat u_0$ is the upper arm and $\hat \ell_0$ is the lower arm. Both arms are written to and read from identical optical memories. \textcolor{black}{We denote the upper and lower memories by $\mathrm{QM}_U$ and $\mathrm{QM}_L$, respectively. Experimentally, each interrogation would consist of writing $\hat u_0$ into $\mathrm{QM}_U$ and $\hat \ell_0$ into $\mathrm{QM}_L$, storing both modes for the same time $\tau$, translating only $\mathrm{QM}_U$ with average velocity $v$ during storage, and then retrieving and recombining the two fields. The motion changes only the accumulated phase; the memory efficiencies and noise floors are otherwise taken identical.} Let $\phi_u$ and $\phi_\ell$ denote the total phases accumulated by the upper and lower arms between the first and second beam splitters. The signal-carrying part of each retrieved arm is modeled as a lossy Gaussian channel,
\begin{align}
\hat u_1 &= e^{i\phi_u}\Bigl[\sqrt{\eta_{\rm mem}(\tau)}\,\hat u_0
+\sqrt{1-\eta_{\rm mem}(\tau)}\,\hat v_u\Bigr], \nonumber\\
\hat \ell_1 &= e^{i\phi_\ell}\Bigl[\sqrt{\eta_{\rm mem}(\tau)}\,\hat \ell_0
+\sqrt{1-\eta_{\rm mem}(\tau)}\,\hat v_\ell\Bigr],
\label{eq:loss_channel}
\end{align}
where $\hat v_u$ and $\hat v_\ell$ are vacuum loss modes. The two memories are assumed to have the same Gaussian write-store-read efficiency~\cite{Gundogan2013, Ahmadi2025Experiment}
\begin{equation}
\eta_{\rm mem}(\tau)=\eta_0\exp\!\left[-\left(\frac{\tau}{\tau_{\rm mem}}\right)^2\right],
\label{eq:eta_mem}
\end{equation}
where $\eta_0$ is the zero-storage efficiency and $\tau_{\rm mem}$ is the Gaussian decay constant.

Only the phase difference matters for the readout, and, \textcolor{black}{to the first order,} we write
\begin{equation}
\phi\equiv\phi_u-\phi_\ell=\phi_b+\phi_v+\delta\phi,
\label{eq:total_phase}
\end{equation}
where $\phi_b$ is the bias phase that sets the operating point, $\phi_v$ is the velocity-dependent signal phase from Eq.~\eqref{eq:phi_v}, and $\delta\phi$ is the residual phase fluctuation. The upper memory \textcolor{black}{$\mathrm{QM}_U$} is the signal arm and the lower memory \textcolor{black}{$\mathrm{QM}_L$} is the reference. \textcolor{black}{Thus $\phi_v$ is the differential phase generated by the motion of $\mathrm{QM}_U$ relative to the stationary reference memory $\mathrm{QM}_L$.}

After the second beam splitter the dark output mode is
\begin{equation}
\hat d=\frac{\hat u_1-\hat \ell_1}{\sqrt{2}}.
\label{eq:dark_output}
\end{equation}
Linearizing around the dark fringe, $\phi_b=0$ and $|\phi_v+\delta\phi|\ll1$, gives
\begin{equation}
\hat d\simeq
\sqrt{\eta(\tau)}\,\hat b_0
+i\sqrt{\eta(\tau)N}\,(\phi_v+\delta\phi)
+\hat d_{\rm vac}
+\hat d_n,
\label{eq:darkport}
\end{equation}
where
\begin{equation}
\eta(\tau)=\eta_{\rm ext}\eta_{\rm mem}(\tau)
=\eta_{\rm ext}\eta_0\exp\!\left[-\left(\frac{\tau}{\tau_{\rm mem}}\right)^2\right]
\label{eq:eta_total}
\end{equation}
is the total signal transmission to the homodyne detector and
\begin{equation}
\eta_{\rm ext}\equiv\eta_{\rm ch}\eta_{\rm det}
\label{eq:eta_ext}
\end{equation}
collects propagation and coupling efficiency $\eta_{\rm ch}$ outside the memories, and detector efficiency $\eta_{\rm det}$. The operator $\hat d_{\rm vac}$ collects vacuum introduced by loss, and $\hat d_n$ is the memory-generated dark-port noise operator defined next.

\subsection{Identical memories and the connection between $\hat d_n$ and $p_n$}

To model unconditional memory noise, let $\hat n_u$ and $\hat n_\ell$ denote the independent phase-insensitive noise fields contributed by the upper and lower memories in the detected temporal mode. Because the two memories are assumed identical, we take
\begin{align}
\avg{\hat n_u} &= \avg{\hat n_\ell}=0, \nonumber\\
\avg{\hat n_u^\dagger \hat n_u} &= \avg{\hat n_\ell^\dagger \hat n_\ell}=p_n(\tau), \nonumber\\
\avg{\hat n_u^\dagger \hat n_\ell} &=0,
\label{eq:identical_noise}
\end{align}
where $p_n(\tau)$ is the detector-referred unconditional memory-noise floor contributed by either memory in one storage trial. The dark-port memory-noise operator is then
\begin{equation}
\hat d_n=\frac{\hat n_u-\hat n_\ell}{\sqrt{2}},
\label{eq:dn_def}
\end{equation}
so that
\begin{equation}
p_n(\tau)=\avg{\hat d_n^\dagger \hat d_n}.
\label{eq:pn_def}
\end{equation}
Thus $\hat d_n$ is the field operator for the dark-port memory noise, while $p_n$ is its mean detector-referred photon number per trial. In the low-noise regime, $p_n$ is equal to the usual unconditional click probability per trial to leading order \cite{Gundogan2015,Heller2022,Esguerra2023,Wang2022,Jutisz2025}.

\subsection{Detection model and full variance}

We assume balanced homodyne detection of the phase quadrature of the dark port,
\begin{equation}
\hat P_d=\frac{\hat d-\hat d^\dagger}{i}.
\label{eq:homodyne_quadrature}
\end{equation}
Residual squeezing-angle jitter rotates the squeezing ellipse by a random angle $\theta$ with root-mean-square width $\sigma_\theta$. Averaging over that rotation gives the effective measured squeezed-quadrature variance
\begin{equation}
V_\theta(r,\sigma_\theta)=\cosh(2r)-e^{-2\sigma_\theta^2}\sinh(2r),
\label{eq:Vtheta}
\end{equation}
which reduces to $e^{-2r}$ when $\sigma_\theta=0$.

The homodyne outcome $x$ for one interrogation is Gaussian,
\begin{equation}
p(x|v)=\frac{1}{\sqrt{2\pi V(\tau)}}
\exp\!\left[-\frac{(x-\mu(v))^2}{2V(\tau)}\right],
\label{eq:likelihood}
\end{equation}
with mean
\begin{equation}
\mu(v)=\sqrt{\eta(\tau)N}\,k_p\tau v
\label{eq:mu_v}
\end{equation}
and variance
\begin{align}
V_{\rm sq}(\tau)
&=1+\eta(\tau)\bigl[V_\theta(r,\sigma_\theta)-1\bigr]+2p_n(\tau) \nonumber\\
&\quad +v_{\rm el}+\eta(\tau)N\sigma_\phi^2.
\label{eq:Vsq}
\end{align}
Here $v_{\rm el}$ is the detector-referred electronic variance in shot-noise units and $\sigma_\phi$ is the root-mean-square residual phase noise per trial. Equation~\eqref{eq:Vsq} contains five contributions: the vacuum unit, the detected squeezing after loss, the unconditional memory photons $p_n$, the electronic noise $v_{\rm el}$, and the phase-noise term $\eta(\tau)N\sigma_\phi^2$. A concise derivation is given in Appendix~\ref{app:variance}.

The coherent-state homodyne benchmark is obtained by setting $r=0$,
\begin{equation}
V_{\rm coh}(\tau)=1+2p_n(\tau)+v_{\rm el}+\eta(\tau)N\sigma_\phi^2,
\label{eq:Vcoh}
\end{equation}
while a classical heterodyne benchmark carries the standard extra vacuum unit,
$V_{\rm het}(\tau)=V_{\rm coh}(\tau)+1$.

\section{Results}

\subsection{Velocity sensitivity and minimum transmission}

Because the variance in Eq.~\eqref{eq:likelihood} is parameter independent in the small-signal regime, the classical Fisher information for one experimental run is
\begin{equation}
F_v^{(1)}(\tau)=\frac{[\partial_v\mu(v)]^2}{V(\tau)}
=\frac{\eta(\tau)N(k_p\tau)^2}{V(\tau)}.
\label{eq:Fv_single}
\end{equation}
For $M$ statistically independent runs,
\begin{equation}
F_v(\tau)=MF_v^{(1)}(\tau),
\label{eq:Fv_total}
\end{equation}
and the Cram\'er-Rao bound is
\begin{equation}
\Delta v(\tau)\ge \frac{\sqrt{V(\tau)}}{k_p\tau\sqrt{M\eta(\tau)N}}.
\label{eq:delta_v}
\end{equation}
In the present Gaussian model the sample mean attains Eq.~\eqref{eq:delta_v}, so the comparisons below are comparisons between achievable Cram\'er-Rao bounds under equal optical resources \cite{Braunstein1994,Paris2009}. Whenever the squeezed-input scheme is favorable, it lowers this bound relative to coherent homodyne with the same $M$ and $N$; it does not, of course, circumvent the bound associated with its own likelihood.

For storage time comparisons at fixed shot number, it is convenient to remove the trivial $1/\sqrt{M}$ scaling and define the shot-normalized sensitivity
\begin{equation}
\Delta v(\tau)\sqrt{M}=\frac{\sqrt{V(\tau)}}{k_p\tau\sqrt{\eta(\tau)N}}.
\label{eq:delta_v_sqrtM}
\end{equation}
This isolates the storage time dependence of the memory-assisted interferometer from the overall number of experimental runs.

The primary classical benchmark is the coherent-state Mach-Zehnder sensor read out by the same homodyne detector, Eq.~\eqref{eq:Vcoh}. It uses the same interferometer, probe resource, and bandwidth. A convenient relative figure of merit is therefore the ratio
\begin{equation}
R(\tau)\equiv\frac{\Delta v_{\rm sq}(\tau)}{\Delta v_{\rm coh}(\tau)}
=\sqrt{\frac{V_{\rm sq}(\tau)}{V_{\rm coh}(\tau)}}.
\label{eq:Rratio}
\end{equation}
The squeezed scheme improves sensitivity only when $R<1$. In the quantum noise-limited regime, the same ratio also gives the required probe power reduction, $P_{\rm sq}/P_{\rm coh}=R^2$.

If $p_n=v_{\rm el}=\sigma_\phi=0$, Eqs.~\eqref{eq:Vsq} and \eqref{eq:Rratio} reduce to the usual pure-loss result,
\begin{equation}
R_{\rm loss}(\tau)=\sqrt{1-\eta(\tau)\bigl[1-V_\theta(r,\sigma_\theta)\bigr]}.
\label{eq:R_loss}
\end{equation}
For the full model with storage-independent $p_n$,
\begin{equation}
R^2(\tau)=1-
\frac{\eta(\tau)\bigl[1-V_\theta(r,\sigma_\theta)\bigr]}
{1+2p_n+v_{\rm el}+\eta(\tau)N\sigma_\phi^2}.
\label{eq:R_full}
\end{equation}
Demanding a target fractional gain $A\equiv1-R$ yields the minimum transmission required for that gain,
\begin{align}
\eta_{\min}(A)
&=\frac{(2A-A^2)(1+2p_n+v_{\rm el})}
{1-V_\theta(r,\sigma_\theta)-(2A-A^2)N\sigma_\phi^2},
\label{eq:eta_min}
\end{align}
provided the denominator is positive.

\textcolor{black}{Unless stated otherwise, the numerical examples use the reference working point summarized in Table~\ref{tab:working_point}. These values are intended as an illustrative near-term operating point rather than as a platform-specific optimum. The photon number in the table corresponds to an energy of $2.5~\mathrm{pJ}$ per trial at $\lambda_p=795~\mathrm{nm}$. For pulse durations between $0.1$ and $1~\mu\mathrm{s}$, this gives average probe powers between roughly $25$ and $2.5~\mu\mathrm{W}$ during the pulse. We use this scale only to set the absolute vertical axis in Fig.~\ref{fig:storage}; whether the weak-probe approximation is satisfied in a specific EIT implementation still depends on beam area, optical depth, and control-field intensity. The memory-noise value in Table~\ref{tab:working_point} is an illustrative detector-referred warm-vapor value consistent with Ref.~\cite{Esguerra2023}, while Refs.~\cite{Gundogan2015,Heller2022} report substantially smaller unconditional noise floors.}

\begin{figure}[t]
    \centering
    \includegraphics[width=\columnwidth]{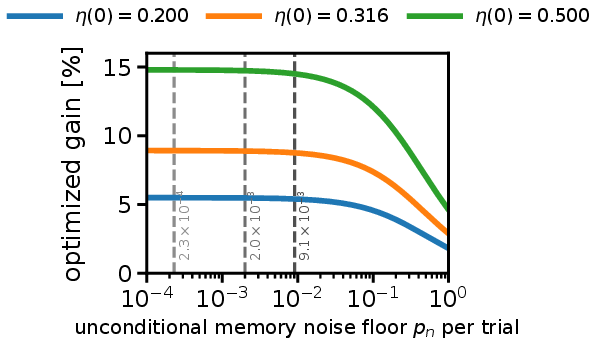}
    \caption{Optimized equal-resource gain versus unconditional memory-noise floor $p_n$ for $10$~dB injected squeezing. Each curve is obtained after numerical optimization of Eq.~\eqref{eq:delta_v} over storage time at fixed shot number for a fixed zero-storage transmission $\eta(0)$. The vertical lines mark representative detector-referred noise floors reported or implied by Refs.~\cite{Heller2022,Gundogan2015,Esguerra2023}. Over the experimentally relevant range $p_n\sim10^{-4}$--$10^{-2}$, and even up to $p_n\sim10^{-1}$, the gain changes only slowly; in the reference working point the advantage is therefore limited more by transmission than by $p_n$.}
    \label{fig:pn_single}
\end{figure}

\begin{table*}[t]
\caption{Reference numerical working point used in the examples. The total
zero-storage transmission is $\eta(0)=\eta_{\rm ext}\eta_0$.}
\label{tab:working_point}
\begin{ruledtabular}
\begin{tabular}{lll}
Parameter & Value & Meaning \\
\hline
$\eta_{\rm ch}$ & $0.67$ & propagation and coupling efficiency outside the memories \\
$\eta_{\rm det}$ & $0.95$ & homodyne detector efficiency \\
$\eta_0$ & $0.50$ & zero-storage write-store-read memory efficiency \\
$\eta_{\rm ext}$ & $0.632$ & external transmission, $\eta_{\rm ch}\eta_{\rm det}$ \\
$\eta(0)$ & $0.316$ & zero-storage total signal transmission, $\eta_{\rm ext}\eta_0$ \\
$p_n$ & $9.1\times10^{-3}$ & detector-referred unconditional memory noise per trial \\
$v_{\rm el}$ & $0.02$ & detector-referred electronic variance in shot-noise units \\
$\sigma_\theta$ & $30~\mathrm{mrad}$ & root-mean-square squeezing-angle jitter \\
$\sigma_\phi$ & $10^{-4}~\mathrm{rad}$ & root-mean-square residual phase noise per trial \\
$N$ & $10^7$ & coherent probe photons per storage trial \\
$\lambda_p$ & $795~\mathrm{nm}$ & probe wavelength \\
$E_{\rm pulse}$ & $2.5~\mathrm{pJ}$ & pulse energy corresponding to $N=10^7$ at $795~\mathrm{nm}$ \\
$T_{\rm pulse}$ & $0.1$--$1~\mu\mathrm{s}$ & representative pulse duration range \\
$P_{\rm pulse}$ & $25$--$2.5~\mu\mathrm{W}$ & average power during the pulse over this duration range \\
\end{tabular}
\end{ruledtabular}
\end{table*}
Figure~\ref{fig:pn_single} summarizes the role of $p_n$ with the other parameters fixed. Even after optimization over storage time, the relative gain decreases only slowly throughout the realistic low-noise range. For $10$~dB injected squeezing, $\eta(0)=0.316$, and the reference $p_n=9.1\times10^{-3}$, the optimized improvement over coherent homodyne is about $5\%$. At the same working point, Eq.~\eqref{eq:eta_min} gives
\begin{align}
\eta_{\min}(10\%)&\approx0.226, \nonumber\\
\eta_{\min}(20\%)&\approx0.437.
\label{eq:eta_numbers}
\end{align}
Inverting Eq.~\eqref{eq:R_full} at fixed transmission then yields an admissible memory-noise budget of order $10^{-1}$ for a $10\%$ gain. For the present parameters, total transmission and phase stability therefore constrain the relative quantum gain more strongly than the unconditional memory floor itself.

\subsection{Optimum storage time for Gaussian lifetime}

Storage time enters in two opposite ways: the signal phase grows with $\tau$, whereas the Gaussian memory lifetime reduces $\eta(\tau)$. For constant $p_n$ and negligible phase noise this tradeoff can be written compactly by introducing the dimensionless time
\begin{equation}
x\equiv\tau/\tau_{\rm mem},
\label{eq:xdef}
\end{equation}
the zero-storage total transmission
\begin{equation}
\eta_{\rm opt}\equiv\eta(0),
\label{eq:etastar}
\end{equation}
the ordinary added-noise floor
\begin{equation}
A_0\equiv1+2p_n+v_{\rm el},
\label{eq:A0}
\end{equation}
and the detected squeezing contrast
\begin{equation}
C\equiv1-V_\theta(r,\sigma_\theta).
\label{eq:Cdef}
\end{equation}
At fixed $M$ and $N$, Eq.~\eqref{eq:delta_v} becomes
\begin{equation}
\Delta v_{\rm sq}(x)=
\frac{\sqrt{A_0}}{k_p\tau_{\rm mem}\sqrt{MN\eta_{\rm opt}}}
\frac{\sqrt{e^{x^2}-c_{\rm eff}}}{x},
\label{eq:delta_v_ceff}
\end{equation}
where
\begin{equation}
c_{\rm eff}\equiv\frac{\eta_{\rm opt}C}{A_0}.
\label{eq:ceff}
\end{equation}
The dimensionless parameter $c_{\rm eff}$ collects the detected squeezing benefit relative to the ordinary added-noise floor. In the fixed-shot model the optimum follows from $\partial_x\Delta v_{\rm sq}=0$ and is
\begin{equation}
\frac{\tau_{\rm opt}}{\tau_{\rm mem}}=
\sqrt{1+W\!\left(-\frac{c_{\rm eff}}{e}\right)},
\label{eq:tau_star_ceff}
\end{equation}
where $W$ is the Lambert-$W$ function and $e$ is Euler's number. In the coherent limit, $V_\theta\to1$ and $c_{\rm eff}\to0$, one recovers $\tau_{\rm opt}=\tau_{\rm mem}$. Increasing $p_n$ moves the optimum back toward this coherent value, while stronger detected squeezing shifts it to shorter storage times.

\begin{figure}[t]
    \centering
    \includegraphics[width=\columnwidth]{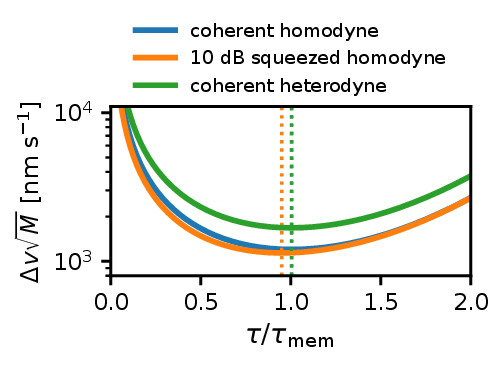}
    \caption{Storage time dependence of the shot-normalized sensitivity $\Delta v\sqrt{M}$ from Eq.~\eqref{eq:delta_v_sqrtM} for coherent homodyne, $10$~dB squeezed homodyne, and coherent heterodyne at the reference $p_n=9.1\times10^{-3}$. The vertical axis is logarithmic and the dotted lines mark the numerical optima. Longer storage initially improves the sensitivity because the signal phase grows with $\tau$, but the accompanying loss of retrieval efficiency becomes dominant at larger $\tau$ and eventually cancels that advantage.}
    \label{fig:storage}
\end{figure}

Figure~\ref{fig:storage} shows the resulting shot-normalized sensitivity on a logarithmic vertical scale. For $\tau_{\rm mem}=100~\mu\mathrm{s}$, the numerical optima are
\begin{align}
\Delta v_{\rm coh}^{\rm opt}\sqrt{M} &\approx 1.20\times10^3~\mathrm{nm~s^{-1}}, \nonumber\\
\Delta v_{\rm sq}^{\rm opt}\sqrt{M} &\approx 1.14\times10^3~\mathrm{nm~s^{-1}}, \nonumber\\
\Delta v_{\rm het}^{\rm opt}\sqrt{M} &\approx 1.68\times10^3~\mathrm{nm~s^{-1}},
\label{eq:opt_numbers}
\end{align}
with optimal storage times near $1.01\tau_{\rm mem}$ for coherent homodyne and $0.95\tau_{\rm mem}$ for squeezed homodyne. The squeezed improvement is real but modest, about $5\%$ relative to coherent homodyne at the same optical resource and shot number.

\subsection{Dependence on squeezing level}

\begin{figure}[t]
    \centering

    \includegraphics[width=\columnwidth]{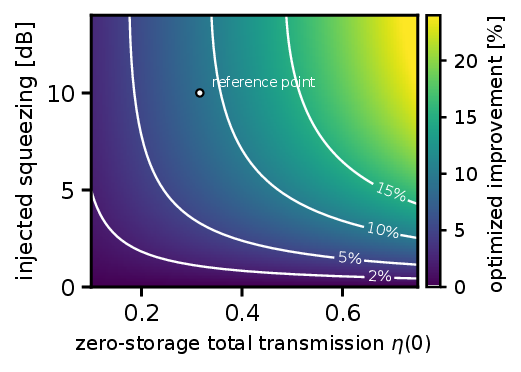}
    \caption{Best equal-resource gain after optimizing Eq.~\eqref{eq:delta_v} over storage time at fixed shot number, shown as a function of the zero-storage total transmission $\eta(0)$ and the injected squeezing level. The reference point used elsewhere in the paper is marked. At fixed transmission the gain saturates once loss and technical noise dominate the detected variance.}
    \label{fig:squeezing_map}
\end{figure}

The injected squeezing level enters through $V_\theta(r,\sigma_\theta)$ and therefore through both the threshold condition in Eq.~\eqref{eq:eta_min} and the effective parameter $c_{\rm eff}$ in Eq.~\eqref{eq:ceff}. Figure~\ref{fig:squeezing_map} maps the best equal-resource gain after numerical optimization over storage time at fixed shot number as a function of the zero-storage total transmission $\eta(0)$ and the injected squeezing level. The map shows three trends. First, increasing squeezing helps only when enough transmission is preserved for that squeezing to reach the detector. Second, at fixed transmission the gain saturates once loss and technical noise dominate the detected variance. Third, the reference point $\eta(0)=0.316$ and $10$~dB injected squeezing lies in a realistic but not especially favorable region: the optimized gain is nonzero, though only at the few-to-ten-percent level.

\section{Discussion}

The model keeps separate the three noise classes that matter experimentally: residual detected squeezing after loss, unconditional memory noise measured with the signal blocked, and readout noise from the interferometer and the detector. This separation is useful when interpreting Figs.~\ref{fig:pn_single}--\ref{fig:squeezing_map}. In the linear-response regime the homodyne record is Gaussian with parameter-independent variance, so the sample mean reaches the Cram\'er-Rao bound of Eq.~\eqref{eq:delta_v}. The plotted gains should therefore be read as regions where the attainable bound for the squeezed-input interferometer falls below the corresponding bound for coherent homodyne under the same optical resource and storage conditions.

Figure~\ref{fig:storage} shows how the signal phase and the retrieval efficiency compete. At short storage times the accumulated phase is too small. At longer times the Gaussian decay of $\eta_{\rm mem}(\tau)$ suppresses the retrieved field strongly enough that the sensitivity rises again. In the constant-$p_n$ model this gives an optimum close to $\tau_{\rm mem}$ for coherent homodyne and a slightly shorter optimum for squeezed input.

Within the reference working point, presently reported low-noise memory floors are already small enough that transmission and residual phase noise are the tighter restrictions on the quantum gain. If those parameters improve, the unconditional memory floor becomes the next quantity that limits how much of the injected squeezing remains useful at the detector. The same balance between experimental run-time gain and storage-induced degradation is expected more broadly in memory-assisted interferometers, including the gravitationally motivated settings discussed in Refs.~\cite{Barzel2024,Gundogan2026Gravity}.

\section{Conclusion}

We have studied how far squeezed light velocimetry can be pushed when the light
is stored and retrieved from realistic quantum memories. In the scheme considered
here, one arm of a two-memory Mach--Zehnder interferometer contains the memory
whose velocity is to be measured. Its translation during storage gives the phase
$\phi_v=-k_p v\tau$. We therefore treat the memory not only as a source of loss,
but as the element that converts motion into an interferometric phase.

The main limitation comes from the same storage time that provides the signal. A
longer storage time increases the velocity-induced phase, but it also reduces the
retrieved transmission and makes the measurement more sensitive to phase noise
and squeezing-angle errors. We find that the best operating point is therefore at
a finite storage time, where phase accumulation and memory degradation are
balanced. For representative near-term parameters, the improvement over coherent
homodyne detection is modest, typically a few to ten percent after optimization,
but it survives realistic optical loss, detector inefficiency, phase noise,
electronic noise, and memory noise.

We also find that the unconditional memory noise levels reported in present experiments, roughly $p_n\sim10^{-4}$--$10^{-2}$ per trial, do not by themselves rule out a squeezed advantage. In the parameter range studied here, total transmission and interferometric phase stability are usually more restrictive. This gives a useful experimental message: reducing incoherent memory noise is important, but the larger gains are likely to come from preserving the squeezed quadrature through the full write--store--read cycle and keeping the interferometer phase stable enough to make use of it. In this sense, the scheme provides a concrete step towards memory-enhanced quantum metrology. The storage time of the memory becomes part of the sensing mechanism, while the same analysis makes clear what must be improved to go beyond the coherent-state shot-noise limit in practice. Higher total efficiency, better phase stability, and faithful preservation of nonclassical quadrature noise are the key requirements.

\textcolor{black}{Although the numerical examples used here are closest to warm-vapor implementations of stored-light velocimetry and related vapor-cell velocity sensors \cite{Ahmadi2025Concept,Ahmadi2025Experiment, Chen2020QEV}, the mechanism is not tied to warm vapors. Cold-atom memories provide another natural platform, with slow-light light-dragging velocimetry and controlled transport of stored light already demonstrated \cite{Kuan2016,Li2020}. Rare-earth-ion-doped memories are also relevant, in particular for long-lived and transportable coherent storage of light \cite{Zhou2022}. In each case, the quantitative advantage is set by the same parameters used in the present model: total efficiency, storage lifetime, phase stability, and preservation of the squeezed quadrature.}

The model can also be used to test related memory-based routes to nonclassical
probe fields. Recent proposals for generating squeezed states via storage in
quantum memories could also be incorporated into this framework
~\cite{Sevincli2026}. More broadly, the results give practical conditions under
which quantum memories can provide a measurable metrological advantage, rather
than only acting as passive storage devices.

\section*{Acknowledgments}
We acknowledge support from the Deutsche Forschungsgemeinschaft (DFG) under project number 448245255 and support from DLR through funds provided by BMFTR (50WM2347, 50SI2655A). MG further acknowledges support from Einstein Foundation Berlin through an Independent Researcher Grant.

\appendix
\setcounter{figure}{0}
\renewcommand{\thefigure}{A\arabic{figure}}
\renewcommand{\theHfigure}{A\arabic{figure}}

\section{Variance model and the connection between $\hat d_n$ and $p_n$}
\label{app:variance}

This appendix shows how Eq.~\eqref{eq:Vsq} is assembled from the dark-port field operator in Eq.~\eqref{eq:darkport}. The result is a variance budget rather than a microscopic Hamiltonian derivation, but each term has a clear experimental meaning.

The dark-port memory-noise operator is defined by Eqs.~\eqref{eq:dn_def} and \eqref{eq:pn_def}. If the memory-generated noise is phase insensitive and diagonal in the number basis,
\begin{equation}
\rho_n(\tau)=\sum_{m=0}^{\infty} P_m(\tau)\,|m\rangle\langle m|,
\qquad
\sum_{m=0}^{\infty} P_m(\tau)=1,
\label{eq:rho_n}
\end{equation}
with mean occupancy
\begin{equation}
p_n(\tau)=\sum_{m=0}^{\infty} mP_m(\tau)=\avg{\hat d_n^\dagger \hat d_n},
\label{eq:pn_appendix}
\end{equation}
then $\avg{\hat d_n}=\avg{\hat d_n^2}=0$ and the quadrature variance is
\begin{equation}
\Var\!\left(\frac{\hat d_n-\hat d_n^\dagger}{i}\right)=1+2p_n(\tau).
\label{eq:pn_to_var}
\end{equation}
Thus the unconditional memory noise floor contributes an excess quadrature variance $2p_n(\tau)$ above vacuum, independent of whether the photon-number distribution is thermal, Poissonian, or otherwise number diagonal.

The squeezed contribution is obtained by propagating the input quadrature variance through loss. With squeezing-angle jitter of root-mean-square width $\sigma_\theta$, the detected squeezed-quadrature variance before external loss is Eq.~\eqref{eq:Vtheta}. A phase-insensitive pure-loss channel with transmission $\eta(\tau)$ transforms any quadrature variance $V$ according to
\begin{equation}
V\longrightarrow \eta(\tau)V+\bigl[1-\eta(\tau)\bigr].
\label{eq:loss_rule}
\end{equation}
Starting from $V_\theta(r,\sigma_\theta)$ therefore gives
\begin{equation}
V_{\rm loss}(\tau)=1+\eta(\tau)\bigl[V_\theta(r,\sigma_\theta)-1\bigr].
\label{eq:Vloss}
\end{equation}

The residual phase noise enters because the dark-port homodyne signal is linear in phase near the operating point. A small phase fluctuation $\delta\phi$ produces an output fluctuation
\begin{equation}
\delta x \simeq \sqrt{\eta(\tau)N}\,\delta\phi,
\label{eq:dx_phase}
\end{equation}
and hence the additive variance
\begin{equation}
\Var(\delta x)=\eta(\tau)N\sigma_\phi^2,
\label{eq:Vphase}
\end{equation}
where $\sigma_\phi^2=\avg{\delta\phi^2}$. Adding also the detector-referred electronic variance $v_{\rm el}$ yields
\begin{align}
V_{\rm sq}(\tau)
&=1+\eta(\tau)\bigl[V_\theta(r,\sigma_\theta)-1\bigr]+2p_n(\tau) \nonumber\\
&\quad +v_{\rm el}+\eta(\tau)N\sigma_\phi^2,
\label{eq:Vsq_appendix}
\end{align}
which is Eq.~\eqref{eq:Vsq} of the main text.

\section{Analytic optimum storage time and the role of $c_{\rm eff}$}

\begin{figure*}[t]
    \centering

    \includegraphics[]{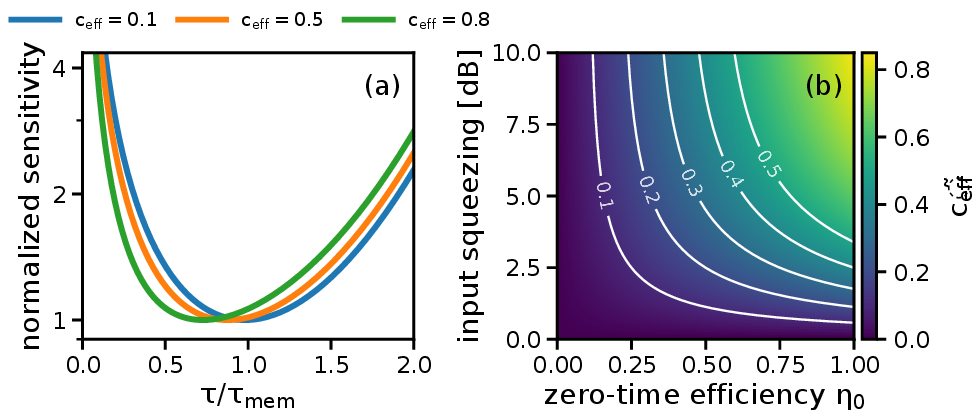}
    \caption{\textbf{(a)} Normalized fixed-shot sensitivity from Eq.~\eqref{eq:ceff_shape} for selected values of $c_{\rm eff}$. The vertical axis is logarithmic. In the analytic fixed-shot model, $c_{\rm eff}$ is the single dimensionless parameter that controls how strongly the optimum is shifted away from the coherent result $\tau_{\rm opt}=\tau_{\rm mem}$. Note that the y-axis is in logarithmic scale. \textbf{(b)} Auxiliary map of the effective parameter $\tilde c_{\rm eff}$ from Eq.~\eqref{eq:ceff_tilde}, plotted versus zero-time memory efficiency $\eta_0$ and injected squeezing at fixed $p_n=0.01$, $N\sigma_\phi^2=0.02$, and $v_{\rm el}=0.02$. The contours show that $\tilde c_{\rm eff}$ grows monotonically with both efficiency and squeezing, but the increase along the squeezing axis gradually weakens once loss and technical noise dominate.}
    \label{fig:appendix_ceff_combined}
\end{figure*}

For fixed $M$ and $N$, negligible phase noise, and storage-independent $p_n$, Eq.~\eqref{eq:delta_v_ceff} may be written in the compact form
\begin{equation}
\Delta v_{\rm sq}(x)\propto \frac{\sqrt{e^{x^2}-c_{\rm eff}}}{x}.
\label{eq:ceff_shape}
\end{equation}
Minimizing the square of this expression gives
\begin{equation}
(1-x^2)e^{x^2}=c_{\rm eff},
\label{eq:ceff_condition}
\end{equation}
whose solution is Eq.~\eqref{eq:tau_star_ceff}. The coherent limit is recovered at $c_{\rm eff}=0$. Figure~\ref{fig:appendix_ceff_combined}a shows the normalized fixed-shot sensitivity landscape for selected values of $c_{\rm eff}$. Small $c_{\rm eff}$ leaves the optimum close to $\tau_{\rm mem}$, while larger $c_{\rm eff}$ shifts the optimum toward shorter storage times.

Figure~\ref{fig:appendix_ceff_combined}b complements Fig.~\ref{fig:appendix_ceff_combined}a by showing how the same parameter varies with zero-time memory efficiency and injected squeezing. For this auxiliary map we keep $p_n$, $v_{\rm el}$, and the phase-noise contribution $N\sigma_\phi^2$ fixed and use the diagnostic combination
\begin{equation}
\tilde c_{\rm eff}=\frac{\eta_0\bigl[1-V_\theta(r,\sigma_\theta)-N\sigma_\phi^2\bigr]}{1+2p_n+v_{\rm el}},
\label{eq:ceff_tilde}
\end{equation}
which reduces to Eq.~\eqref{eq:ceff} when phase noise is neglected and external losses are absorbed into the horizontal efficiency axis. The map makes it clear that sizeable shifts of the optimum storage time require both reasonably high zero-time efficiency and several decibels of useful squeezing.

\bibliographystyle{apsrev4-2}
\bibliography{references}

\end{document}